\documentclass[pra,twocolumn,superscriptaddress]{revtex4-1}
\usepackage{amssymb}
\usepackage{graphicx}
\usepackage{dcolumn}
\usepackage{bm}
\usepackage{amsmath}
\usepackage[usenames]{color}
\usepackage{textcomp}
\usepackage{subfigure}
\usepackage{comment}
\allowdisplaybreaks[4]

\usepackage[colorlinks,linkcolor=blue,anchorcolor=blue,citecolor=blue,urlcolor=blue,driverfallback=dvipdfm]{hyperref}

\begin{document}

\title{$p$-orbital self-organization of ultracold atoms
coupled to optical cavities}
\author{Hui Tan}
\affiliation{Department of Physics, National University of Defense Technology, Changsha 410073, P. R. China}
\author{Pengfei Zhang}
\affiliation{Department of Physics, National University of Defense Technology, Changsha 410073, P. R. China}
\author{Jianmin Yuan}
\affiliation{Institute of Atomic and Molecular Physics, Jilin University, Changchun 130012, P. R. China}
\affiliation{Department of Physics, National University of Defense Technology, Changsha 410073, P. R. China}
\author{Yongqiang Li}
\email{li\_yq@nudt.edu.cn}
\affiliation{Department of Physics, National University of Defense Technology, Changsha 410073, P. R. China}
\affiliation{Hunan Key Laboratory of Extreme Matter and Applications, National University of Defense Technology, Changsha 410073, P. R. China}
 \affiliation{Hunan Research Center of the Basic Discipline for Physical States, National University of Defense Technology, Changsha 410073, China}

\begin{abstract}
Atoms coupled to optical cavities provide a novel platform for understanding high-orbital exotic phenomena in strongly correlated materials. In this study, we investigate strongly correlated ultracold bosonic gases that are coupled to two orthogonally arranged optical cavities and driven by a blue-detuned running-wave laser field. Our results demonstrate that atoms initially in the $s$-orbital state can be scattered into $p_x$- and $p_y$-orbital states in either a symmetric or asymmetric manner, depending on the frequencies of the two cavities. For the symmetric configuration, we observe that atoms are scattered into the $p_x$- and $p_y$-orbitals equally. In the asymmetric case, photons emitted into one cavity mode suppress the scattering into the orthogonal mode. Notably, the coupling of atoms with multiple cavity modes leads to the emergence of high-orbital self-organized phases, accompanied by orbital-density waves that break different symmetries. 

\end{abstract}

\date{\today}
\maketitle


\section{\label{sec:level1}introduction}
Cavity-BEC systems serve as a versatile testbed for quantum many-body physics, leveraging cavity photons to engineer tunable, infinite-range interactions between atoms. This platform facilitates studies of superradiant phase transitions, self-organization dynamics, and measurement-induced entanglement in driven-dissipative settings. In recent years, extensive theoretical~\cite{PhysRevLett.89.253003, PhysRevLett.110.113606, PhysRevLett.111.055702,PhysRevLett.112.143002,PhysRevLett.112.143003,PhysRevLett.112.143004,PhysRevLett.114.173903,PhysRevA.87.051604,PhysRevLett.114.123601} and experimental~\cite{2010Nature,PhysRevLett.113.070404,2016Nature,PhysRevLett.115.230403,PhysRevLett.120.223602,PhysRevLett.121.163601,norcia2018cavity,kroeze2019dynamical,Mivehvar02012021} investigations have been conducted on the superradiant phase, which is characterized by spontaneous $Z_2$ symmetry breaking through intra-orbital scattering, achieved by coupling atoms to red-detuned optical cavities.  
Furthermore, coupling atoms to two hyperfine cavity modes has established spontaneous $U(1)$ symmetry breaking as a central research focus, supported by experimental~\cite{RN77,10.1126/science.aan2608,s41563-018-0118-1} and theoretical~\cite{Lang_2017,PhysRevA.96.063828} work. Investigations have also been extended to coupled order parameters~\cite{morales_coupling_2018}, higher-order topological Peierls insulators~\cite{PhysRevLett.131.263001}, and subradiance phenomena~\cite{donner2024stabilitydecaysubradiantpatterns}. Most recently, studies leveraging more cavity modes have explored $SO(3)$ and quasicrystalline symmetries~\cite{PhysRevA.98.023617,PhysRevLett.123.210604}.  

While red-detuned pumping has been extensively studied, atoms interacting with optical cavities and driven by blue-detuned pump lasers provide a complementary and promising platform, particularly for exploring high-orbital exotic phenomena in the context of strongly correlated materials. Although it was initially thought that atoms at intensity minima under blue-detuned light would struggle to scatter photons into the cavity mode, both experimental observations and theoretical studies have now demonstrated that superradiant phase transitions can indeed occur in high-finesse cavities pumped by blue-detuned light, as predicted~\cite{PhysRevLett.105.043001,PhysRevA.83.033601,PhysRevA.85.013817,PhysRevLett.115.163601,PhysRevLett.118.073602,PhysRevA.99.053605,Ke_ler_2020,PhysRevA.101.061602,PhysRevResearch.3.013173} and experimentally confirmed~\cite{PhysRevLett.123.233601,PhysRevResearch.3.L012024}. Crucially, in contrast to the charge-density waves characterizing superradiance under red-detuned pumping, the self-organization induced by blue-detuned pumping is associated with orbital flipping caused by inter-orbital scattering~\cite{PhysRevLett.123.233601,PhysRevA.106.023315}. This distinct mechanism offers a novel route to investigate exotic orbital physics, presenting unique properties compared to conventional methods like lattice shaking~\cite{eckardt2017colloquium,bukov2015universal} and bipartite-lattice configurations~\cite{wirth2011evidence,PhysRevLett.106.015302,PhysRevLett.114.115301,PhysRevLett.126.035301,RN64}.
Nevertheless, the combination of high orbitals with multiple optical cavities pumped by a blue detuned laser has not been previously investigated. The orbital anisotropy and long-range interactions resulting from multiple cavities may yield intriguing high-orbital physics.

\begin{figure}[t]
\includegraphics[width=1.0\linewidth]{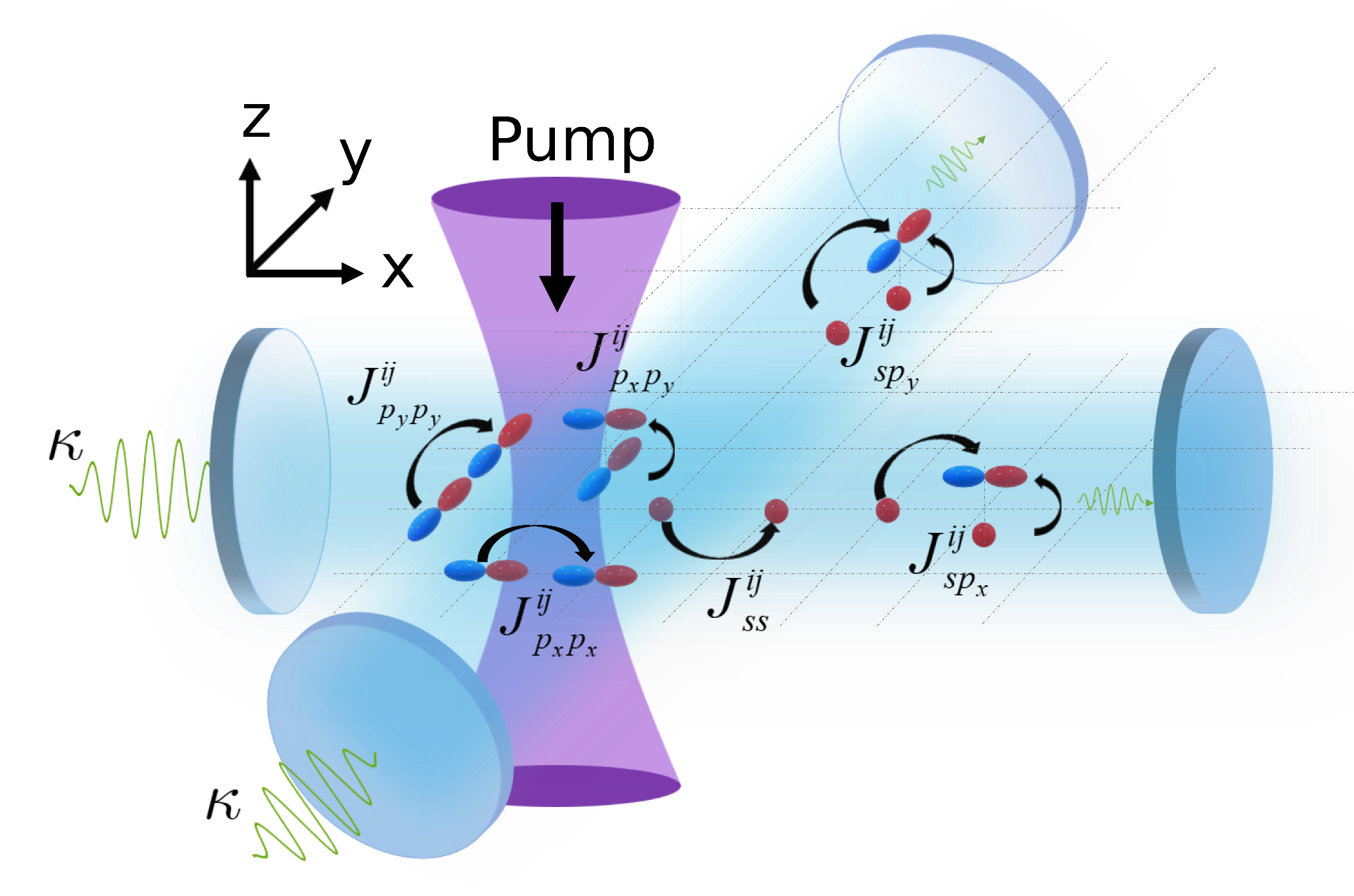}
\caption{$p$-orbital superradiance induced by coupling atoms to two orthogonal high-finesse cavity modes pumped by a blue-detuned running wave laser in the transverse direction. Here, we display the dominant inter-orbital scattering processes, where $J^{ij}_{sp_x}$, $J^{ij}_{sp_y}$, and $J^{ij}_{p_xp_y}$ denote the onsite and nearest-neighbour scattering processes between the $s$-, $p_x$-, and $p_y$-orbital bands. $J^{ij}_{ss}$, $J^{ij}_{p_xp_x}$, and $J^{ij}_{p_yp_y}$ are the normal nearest-neighbour hopping terms for the $s$-, $p_x$-, and $p_y$-orbital states, respectively.}
\label{figure1}
\end{figure}

In this study, we propose an approach to realize $p$-orbital quantum phases in ultracold bosonic gases by coupling atoms to two orthogonal cavities, stimulated by a blue-detuned running wave laser. In this configuration, atoms initially in the $s$-orbital state are scattered into $p_x$- or $p_y$-orbital states, stabilizing the $p$-orbital superradiant phase. Rich phase diagrams are constructed based on the developed three-component real-space bosonic dynamical mean-field theory. For the identical cavity setup, atoms are symmetrically scattered into the $p_x$- and $p_y$-orbitals. However, when the cavity mode frequencies differ, atoms are asymmetrically scattered into the $p_x$- or $p_y$-orbitals, leading to the stabilization of distinct high-orbital many-body phases. These phases are accompanied by orbital-density waves that break different types of symmetries. Our study provides a promising platform for exploring exotic high-orbital many-body physics. 

This paper is organized as follows: Section~\ref{sec:level2} provides an introduction to the extended Bose-Hubbard model and outlines the formulation of the three-component bosonic dynamical mean-field theory. Section~\ref{sec:level3} presents the results concerning the high-orbital superradiant phases under both symmetric and asymmetric cavity configurations. Finally, Section~\ref{sec:level4} concludes the paper with a comprehensive discussion.

\section{\label{sec:level2} Model and Method}
We consider $N$ two-level ultracold atoms with mass $m$ and
transition frequency $\omega_a$ interacting with two single-mode cavities
with frequencies $\omega_{1}$ and $\omega_2$ in the $x$ and $y$ directions, respectively. The atom-cavity system is driven by a
running wave pump field with frequency $\omega_p$ in the $z$ direction, as shown in Fig.~\ref{figure1}. The system can be characterized by an effective Hamiltonian, which is derived from the Jaynes-Cummings Hamiltonian under the rotating wave approximation and dipole approximation. Following the standard procedure outlined in Appendix A, the many-body Hamiltonian can be described by
\begin{eqnarray}\label{Hamil_1}
H&=&\int \mathrm{d}{\bf r} \hat{\Psi}^\dagger ({\bf r})\left[-\frac{\hbar^2\nabla^2}{2m}  + \hat{V}({\bf r}) \right] \hat{\Psi}({\bf r})\\
 &+&\int \mathrm{d}{\bf r} \hat{\Psi}^\dagger ({\bf r})(\hat{V}_{1,\mathrm{scat}} +\hat{V}_{2,\mathrm{scat}} +\hat{V}_{12,\mathrm{scat}} )\hat{\Psi}({\bf r})\nonumber\\
&-&\sum_{\sigma=1,2}\hbar\Delta_\sigma \hat{a}^\dagger_\sigma \hat{a}_\sigma +\frac{g}{2}\int \mathrm{d}{\bf r} \hat{\Psi}^\dagger({\bf r})\hat{\Psi}^\dagger({\bf r})\hat{\Psi}({\bf r})\hat{\Psi}({\bf r})\nonumber,
\end{eqnarray}
where $V({\bf r})=U_0\hat{a}^\dagger_1 \hat{a}_1\sin^2(kx) + U_0\hat{a}^\dagger_2 \hat{a}_2 \sin^2(ky)$ is the standing-wave potential created by the cavities in the $xy$ plane. $\hat{V}_{1,\mathrm{scat}}=\frac{\sqrt{V_pU_0}}{2}\sin(kx)\left(\hat{a}_1+\hat{a}^\dagger_1 \right)$, and $\hat{V}_{2,\mathrm{scat}}=\frac{\sqrt{V_pU_0}}{2}\sin(ky)\left(\hat{a}_2 +\hat{a}^\dagger_2  \right)$ are scattering terms induced by cavities in $x$ and $y$ directions, respectively. $\hat{V}_{12,\mathrm{scat}}=U_0\sin(kx)\sin(ky)\left(\hat{a}^\dagger_1 \hat{a}_2+\hat{a}^\dagger_2 \hat{a}_1 \right)$ is scattering between the two cavities. $\hat{\Psi}({\bf r})$ denotes the atomic field operator for annihilating an atom at position ${\bf r}$ in the ground state, where the excited states have been eliminated for large detuning for both cavities~\cite{PhysRevLett.95.260401,PhysRevLett.100.050401}. $\Delta_1 \equiv \omega_p - \omega_1$, and $\Delta_2 \equiv \omega_p - \omega_2$, $V_p$ is the strength of the pump laser, and $U_0$ is the lattice depth induced by single cavity  photon. $\hat{a}_\sigma^\dagger$ ($\hat{a}_\sigma$) is the photon creation (annihilation) operator for the cavity $\sigma$. Finally, $g=\frac{4\pi\hbar^2a_s}{m}$ describes the contact interaction between atoms, where $a_s$ is the $s$-wave scattering length.

We study the blue-detuned regime with $\omega_p-\omega_a>0$ and $\omega_\sigma-\omega_a>0$, where $p$-orbital superradiance in optical cavities has been experimentally demonstrated using $^{87}$Rb atoms in a high-finesse single-mode cavity~\cite{PhysRevLett.123.233601}. Our model uses experimentally relevant parameters with both cavity modes sharing identical decay rates $\kappa = 40\, \omega_r$ ($\omega_r=\hbar k^2/2m$ being the recoil frequency), and atoms are pumped by a running blue-detuned laser with wavelength $\lambda_p = 780.1\mathrm{nm}$. To realize a two-dimensional lattice system, atomic motion along the $z$ direction is frozen by employing an intense standing-wave laser with a potential depth of $V_z = 50E_r$, where $E_r = \hbar \omega_r$ denotes the recoil energy.

In this study, we primarily focus on the strongly interacting regime. Under conditions of a sufficiently deep lattice potential, the tight-binding model can be rigorously derived. We impose additional lattice potentials $V_{x,y}=5E_r$ in the $x$ and $y$ directions, respectively, to ensure the validity of the tight-binding approximation. By retaining the lowest three bands ($s$, $p_x$, and $p_y$), the many-body system can be effectively described by an extended Bose-Hubbard model (see Appendix A)
\begin{eqnarray}\label{Hamil_2}
\hat{H} &=&  - \sum_{\langle i, j\rangle,\sigma}J^{ij}_{
\sigma\sigma}(\hat{b}^\dagger_{i,\sigma} \hat{b}_{j,\sigma}+ {\rm H.c.}) + \sum_{i,\sigma}\mu_{\sigma}\hat{b}^\dagger_{i,\sigma} \hat{b}_{i,\sigma}\nonumber\\ &+&\sum\limits_{i,\sigma _{1}\sigma _{2}\sigma _{3}\sigma _{4}}\frac{%
U_{\sigma _{1}\sigma _{2}\sigma _{3}\sigma _{4}}}{2}\hat{b}_{i,\sigma
_{1}}^{\dag }\hat{b}_{i,\sigma _{2}}^{\dag }\hat{b}_{i,\sigma _{3}}\hat{b}_{i,\sigma _{4}}\nonumber\\
&+&\hat{V}_1+\hat{V}_2+\hat{V}_{12}-\hbar |\alpha_1|^2 \Delta_1 - \hbar |\alpha_2|^2 \Delta_2, 
\end{eqnarray}
where $\langle i,j\rangle$ denotes summation over nearest-neighbor sites, $J_{\sigma_1 \sigma_2}^{ij}$ represents the onsite (for $i = j$) and nearest-neighbor (for $i \neq j$) single-particle hopping amplitudes, $\mu_\sigma \equiv J^{ii}_{\sigma\sigma}$ is the chemical potential, and $U_{\sigma_1 \sigma_2 \sigma_3 \sigma_4}$ corresponds to the onsite interaction terms. $\hat{b}_{i,\sigma}$ represents the annihilation operator for the Wannier state $\sigma$ at site $i$, where $\sigma$ refers to the $s$-, $p_x$-, and $p_y$-orbitals, respectively.  $\hat{V}_{1}=2\mathrm{Re}[\alpha_1]\sum_{ij} (-1)^{i_x}\left(J^{ij}_{sp_x} \hat{b}^\dagger_{i,s} \hat{b}_{j,p_x} + \mathrm{H.c.} \right)$, and $\hat{V}_{2}=2\mathrm{Re}[\alpha_2]\sum_{ij} (-1)^{i_y}\left(J^{ij}_{sp_y} \hat{b}^\dagger_{i,s} \hat{b}_{j,p_y} + \mathrm{H.c.} \right)$ are the scattering processes between the pump laser and each of the two cavities, respectively. $V_{12}=2\left( \mathrm{Re}[\alpha_1]*\mathrm{Re}[\alpha_2] + \mathrm{Im}[\alpha_1]*\mathrm{Im}[\alpha_2] \right)\sum_{ij} (-1)^{i_x + i_y}\\
\times\left(J^{ij}_{p_xp_y} \hat{b}^\dagger_{i,p_x} \hat{b}_{j,p_y} + \mathrm{H.c.} \right)$ denotes the inter-cavity scattering process. While the cavity modes are macroscopically populated, cavity fields can be simplified by the coherent-state approximation $\alpha_1 (t)=\langle \hat{a}_1(t)\rangle$ and $\alpha_2 (t)=\langle \hat{a}_2(t)\rangle$~\cite{PhysRevLett.107.140402}. We then determine the cavity fields in the steady state for large decay rates $\kappa$, with 
\begin{eqnarray}\label{alpha_1}
\alpha_1 &=&\sum_{i}%
\frac{(-1)^{i_x}J^{ii}_{sp_x}\langle\hat{b}_{i,s}^{\dagger }\hat{b}_{i,p_x}+\mathrm{H.c.}\rangle}{\Delta _{1}-\sum_{i,\sigma
	}\langle J_{1,\sigma }\hat{b}_{i,\sigma }^{\dagger }\hat{b}_{i,\sigma }\rangle
	+i\kappa}\\ 
 &+&\sum_{i}\frac{\alpha_2 (-1)^{i_x+i_y}J^{ii}_{p_{x}p_{y}}\langle\hat{b}%
	_{i,p_x}^{\dagger }\hat{b}_{i,p_{y}} +\mathrm{H.c.}\rangle}{ \Delta _{1}-\sum_{i,\sigma
	}\langle J_{1,\sigma }\hat{b}_{i,\sigma }^{\dagger }\hat{b}_{i,\sigma }\rangle
	+i\kappa },\nonumber
\end{eqnarray}
and 
\begin{eqnarray}\label{alpha_2}
\alpha_2 &=&\sum_{i}%
\frac{(-1)^{i_y}J^{ii}_{sp_y}\langle\hat{b}_{i,s}^{\dagger }\hat{b}_{i,p_y}+\mathrm{H.c.}\rangle}{\Delta _{2}-\sum_{i,\sigma
	}\langle J_{2,\sigma }\hat{b}_{i,\sigma }^{\dagger }\hat{b}_{i,\sigma }\rangle
	+i\kappa}\\ 
 &+&\sum_{i}\frac{\alpha_1 (-1)^{i_x+i_y}J^{ii}_{p_{x}p_{y}}\langle\hat{b}%
	_{i,p_x}^{\dagger }\hat{b}_{i,p_{y}} +\mathrm{H.c.}\rangle}{ \Delta _{2}-\sum_{i,\sigma
	}\langle J_{2,\sigma }\hat{b}_{i,\sigma }^{\dagger }\hat{b}_{i,\sigma }\rangle
	+i\kappa }.\nonumber
\end{eqnarray}
The Hubbard parameters in Eq.~(\ref{Hamil_2}) are determined through self-consistent calculations based on Wannier function analysis of the cavity-induced two-dimensional square lattice (see Appendix A). 

For blue atom-pumping detuning, atoms located at the minimum of the lattice potential~\cite{PhysRevLett.123.233601}, result in the scattering terms $V_{1,\mathrm{scat}}$ and $V_{2,\mathrm{scat}}$ being odd parity in the $x$ and $y$ directions, respectively. And $V_{12,\mathrm{scat}}$ are odd parity in both $x$ and $y$ directions. Thus, atoms are scattered into higher orbitals, in contrast to the case of red atom-pumping detuning, where atoms can only be scattered into the same orbital with higher momenta~\cite{2010Nature}. The scattering processes between different orbitals are shown in Fig.~\ref{figure1}, where $V_{1,\mathrm{scat}}$ scatters the atoms from the $s$- to $p_{x}$-orbital,  $V_{2,\mathrm{scat}}$ scatters the atoms from the $s$- to $p_{y}$-orbital, and scattering between $p_{x}$- and $p_{y}$-orbitals is induced by $V_{12,\mathrm{scat}}$. The scattering between $s$- and $d_{xy}$-orbitals induced by $V_{12,\mathrm{scat}}$ is neglected due to the relatively weak nature of the scattering between cavities.

To compute the equilibrium solution of Eq.~(\ref{Hamil_2}), we implement a three-component bosonic dynamical mean-field theory (BDMFT). BDMFT presents a nonperturbative method for analyzing many-body systems in both three and two dimensions \cite{Vollhardt, Hubener, Werner, Li2011,li2024heteronuclear}, with its accuracy validated through comparisons with quantum Monte Carlo simulations \cite{QMC_boson}. Within the framework of BDMFT, we derive the effective action for the impurity site following the standard way~\cite{Appen_georges96, Vollhardt}
\begin{widetext}
\begin{eqnarray}\label{eff_action}
	S^{(0)}_\mathrm{imp}&=&\int_{0}^{\beta} d\tau d\tau'\sum_{\sigma_1,\sigma_1',\sigma_2, \sigma_2'}\left(
	\begin{array}{c} b^*_{0,\sigma_1}(\tau)\\
		b_{0,\sigma_1}(\tau)
	\end{array}
\right)^{T}\mathcal{G}^{-1}_{0,\sigma_1\sigma_2\sigma_1'\sigma_2'}(\tau-\tau')
	\left(
	\begin{array}{c} b_{0,\sigma_2}(\tau')\\
		b^*_{0,\sigma_2}(\tau')
	\end{array}
	\right)\nonumber\\
 &-&\int_{0}^{\beta} d\tau \sum_{\langle 0j \rangle, \sigma_1,\sigma_1^\prime}(-1)^{\delta_{\sigma_1\sigma_1^\prime}+1}J^{0j}_{\sigma_1\sigma_1^\prime}[b^\ast_{0,\sigma_1}(\tau) \phi_{j,\sigma_1'}(\tau)+{\rm H.c.}]
	\nonumber\\
	&+&\int_{0}^{\beta} d\tau\left( \sum_{\sigma_1,\sigma_1^\prime}J^{00}_{\sigma_1\sigma_1^\prime}b^\ast_{0,\sigma_1}(\tau)b_{0,\sigma_1^\prime}(\tau)+{\rm H.c.}+\frac12\sum_{\sigma_1\sigma_2\sigma_3\sigma_4}U_{\sigma_1\sigma_2\sigma_3\sigma_4}b^{(0)\ast}_{\sigma_1}(\tau)b^{(0)\ast}_{\sigma_2}(\tau)b^{(0)}_{ \sigma_3}(\tau)b^{(0)}_{\sigma_4}(\tau)\right),
\end{eqnarray}
\end{widetext}
where $J^{00}_{\sigma_1\sigma_1'}$ is the cavity induced onsite scattering, and $J^{0j}_{\sigma_1\sigma_1'}$ denotes the nearest-neighbor hopping and scattering terms. The Weiss Green's function is defined as
\begin{widetext}
\begin{eqnarray}
&&\mathcal{G}^{-1}_{0,\sigma_1\sigma_2\sigma_1'\sigma_2'}(\tau-\tau')=\\
	&&\left(
	\begin{array}{cc}
		(\partial_{\tau'}-\mu_{\sigma_1})\delta_{\sigma_1\sigma_2}+\sum_{\langle 0j \rangle,\langle 0j' \rangle}J^{0j}_{\sigma_1\sigma_1'}J^{0j'}_{\sigma_2\sigma_2'}G^1_{j,j',\sigma_1',\sigma_2'}(\tau,\tau') & \sum_{\langle 0j \rangle,\langle 0j' \rangle}J^{0j}_{\sigma_1\sigma_1'}J^{0j'}_{\sigma_2\sigma_2'}G^2_{j,j',\sigma_1',\sigma_2'}(\tau,\tau')\\
		\sum_{\langle 0j \rangle,\langle 0j' \rangle}J^{0j}_{\sigma_1\sigma_1'}J^{0j'}_{\sigma_2\sigma_2'}G^{*2}_{j,j',\sigma_1',\sigma_2'}(\tau',\tau) &(-\partial_{\tau'}-\mu_{\sigma_1})\delta_{\sigma_1\sigma_2}+\sum_{\langle 0j \rangle,\langle 0j' \rangle}J^{0j}_{\sigma_1\sigma_1'}J^{0j'}_{\sigma_2\sigma_2'}G^{*1}_{j,j',\sigma_1',\sigma_2'}(\tau',\tau)
	\end{array}
	\right)\nonumber.
\end{eqnarray}
\end{widetext}
The superfluid order parameter is given by
\begin{equation}
	\phi^{}_{j,\sigma_1}(\tau) \equiv \langle b_{j, \sigma_1} (\tau)
	\rangle_0,
\end{equation}
and the connected Green's functions read
\begin{eqnarray}
	&&G^1_{j,j',\sigma_1',\sigma_2'}(\tau,\tau')\nonumber\\
 &&=\langle b_{j,\sigma_1'}(\tau) b^*_{j',\sigma_2^\prime}(\tau')\rangle_{(0)}-  \phi_{j,\sigma_1'}(\tau) \phi^*_{j',\sigma_2^\prime}(\tau'),
\end{eqnarray}
\begin{eqnarray}
&&G^2_{j,j',\sigma_1',\sigma_2'}(\tau,\tau')\nonumber\\
 &&=\langle b_{j,\sigma_1'}(\tau) b_{j',\sigma_2'}(\tau')\rangle_{(0)}-  \phi_{j,\sigma_1'}(\tau) \phi_{j',\sigma_2'}(\tau'),
\end{eqnarray}
for diagonal and off-diagonal parts, respectively. In the cavity method, $\langle \ldots \rangle_0$ takes the expectation value  excluding the impurity site.

The effective action Eq.~(\ref{eff_action}) can be solved by diagram perturbation theory or the Hamiltonian-based method. In the study here, we adopt the second one and map Eq.~(\ref{eff_action}) to the Anderson impurity Hamiltonian
\begin{eqnarray}
&&\hat{H}^{(0)}_A=- \sum_{j,\sigma}J^{0j}_{\sigma\sigma} \left(\phi^{(j)*}_{\sigma} \hat{b}^{(0)}_{\sigma} + {\rm H.c.} \right)\\
&-&\sum_{\sigma}\mu_{\sigma} \hat{n}^{(0)}_{\sigma}+\frac12\sum_{\sigma_1\sigma_2\sigma_3\sigma_4}U_{\sigma_1\sigma_2\sigma_3\sigma_4}b^{(0)\ast}_{\sigma_1}b^{(0)\ast}_{\sigma_2}b^{(0)}_{ \sigma_3}b^{(0)}_{\sigma_4} \nonumber  \\
&+&2\mathrm{Re}[\alpha_1] \left[J^{00}_{sp_x}\hat{b}_{s}^{(0)*} \hat{b}_{p_x}^{(0)} + \sum_{j}J^{0j}_{sp_x}\hat{b}_{s}^{(0)*} \phi_{p_x}^{(j)} +{\rm H.c.} \right]\nonumber\\
&+& 2\mathrm{Re}[\alpha_2]\left[J^{00}_{sp_y} \hat{b}_{s}^{(0)*} \hat{b}_{p_y}^{(0)} + \sum_{j}J^{0j}_{sp_y} \hat{b}_{s}^{(0)*} \phi_{p_y}^{(j)} +{\rm H.c.}\right]\nonumber\\
&+&2\left( \mathrm{Re}[\alpha_1]*\mathrm{Re}[\alpha_2] + \mathrm{Im}[\alpha_1]*\mathrm{Im}[\alpha_2] \right)\nonumber\\
&\times&\left[J^{00}_{p_xp_y}\hat{b}_{p_x}^{(0)*} \hat{b}_{p_y}^{(0)} + \sum_{j}J^{0j}_{p_xp_y}\hat{b}_{p_x}^{(0)*} \phi_{p_y}^{(j)} +{\rm H.c.} \right]\nonumber\\
&+& \sum_{l}  \epsilon_l \hat{a}^\dagger_l\hat{a}_l + \sum_{l,\sigma} \Big( V_{\sigma,l} \hat{a}^\dagger_l\hat{b}^{(0)}_{\sigma} + W_{\sigma,l} \hat{a}_l\hat{b}^{(0)}_{\sigma} + {\rm H.c.} \Big).\nonumber
\end{eqnarray}
BDMFT simplifies the many-body lattice problem to a single impurity problem. The onsite terms for this impurity site are directly inherited from the Hubbard Hamiltonian, such as chemical potential, contact interaction, and onsite orbital-flip terms.  The coupling between the impurity and its surrounding lattice sites is characterized by two distinct baths: the condensed and normal baths. Where the condensed bath is represented by the superfluid order parameters $\phi^{(0)}_{\sigma}$. The normal bath, characterized by a finite number of orbitals with creation operators $\hat{a}^\dagger_l$ and energies
$\epsilon_l$, is coupled to the impurity site through normal-hopping $V_{\sigma, l}$ and anomalous-hopping amplitudes $W_{\sigma, l}$. Then, the Anderson Hamiltonian can be solved for every lattice site by exact diagonalization under the Fock basis~\cite{Appen_georges96, Hubener}. In our numerical calculation, the total number of lattice sites is chosen as $N_{\mathrm{lat}}=16\times 16$, and the Fock basis is truncated by ($N_s, N_{p_x}, N_{p_y}, N_L, L$), where $N_s$, $N_{p_x}$, and $N_{p_y}$ are truncation numbers for occupation on $s$-, $p_x$- and $p_y$-orbitals, respectively. $L$ is the number of orbitals for the normal bath, and $N_L$ truncates occupation on these orbitals. Appropriate truncations are selected to minimize computational costs while maintaining the accuracy of the results.

\section{\label{sec:level3} Results}
\begin{figure}[tbp]
\includegraphics[width=1.0\linewidth]{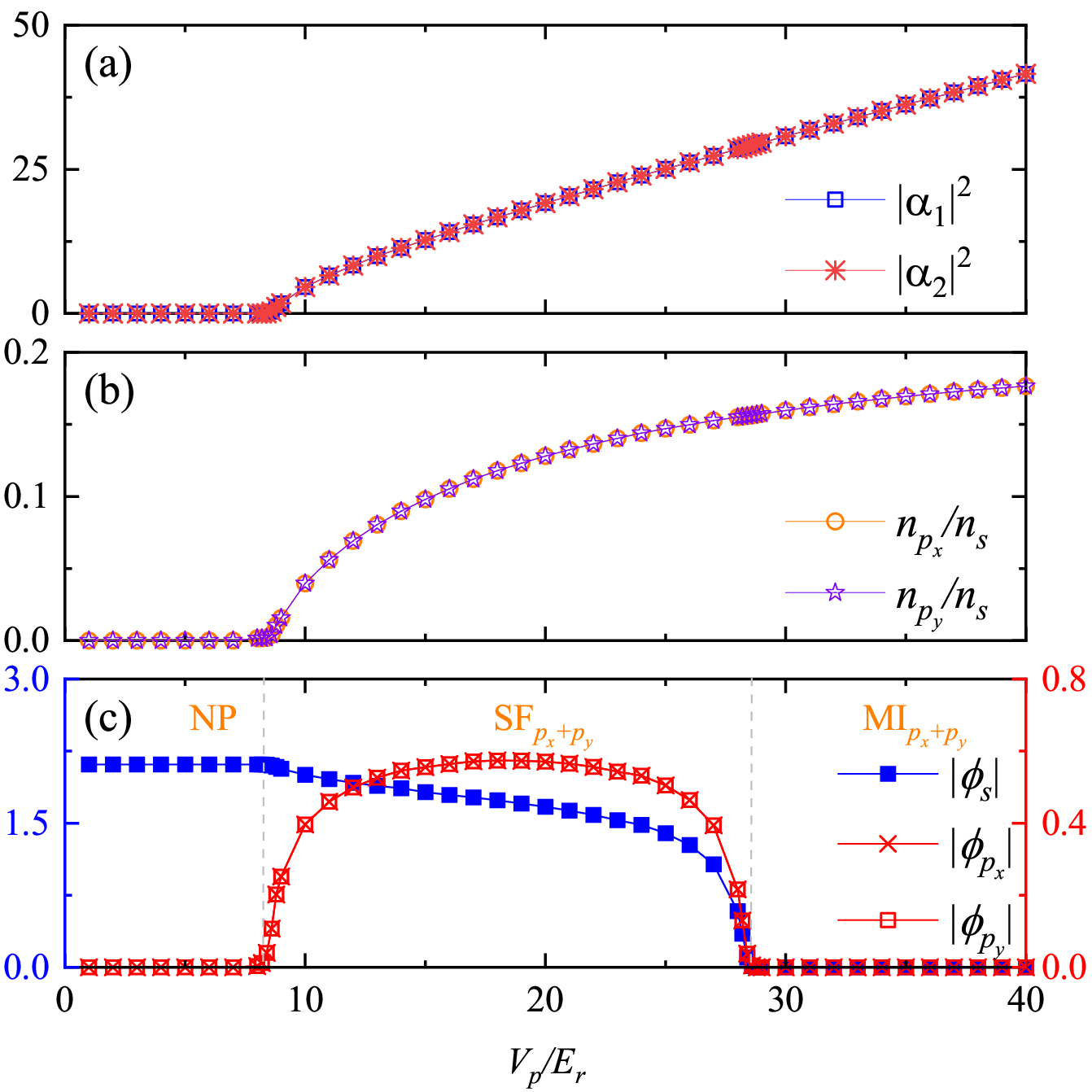}
\caption{Order parameters of ultracold bosonic gases pumped by a blue-detuned laser with the symmetric cavity detuning $\Delta_{1,2}=0$, obtained from bosonic dynamical mean-field theory. (a) Photon numbers in the $x$- and $y$-directional cavities as a function of pump strength $V_p$.  (b) Ratio of atom population for different orbital states. (c) Superfluid order parameters for the $s$-, $p_x$-, and $p_y$-orbital states, respectively. As the pump strength increases, there exist normal (NP), $p$-orbital superfluid (SF$_{p_x+p_y}$), and $p$-orbital Mott-insulating (MI$_{p_x+p_y}$) phases. Here, $N\times U_0=150E_r$ with $N$ being the total number of particles, and decay rate $\protect\kappa=40\, \protect\omega_r$.}
\label{figure2}
\end{figure}

\subsection{\label{sec:level3A} $p$-orbital self-organization for $\Delta_1=\Delta_2$}
We first discuss the physics for the identical cavity configuration $\Delta_1=\Delta_2$, where atoms in $s$-orbital are symmetrically scattered into $p_x$- and $p_y$-orbitals. The steady-state phase diagram is evaluated using distinct order parameters. These encompass the superfluid order parameter $|\phi _{\sigma }|=|\langle \hat{b}_{i,\sigma }\rangle| $, the ratio of atom population on different orbital states $n_{p_x, p_y}/n_s$, where $n_\sigma = \sum_{i}n_{i,\sigma}$ represents the total atoms in the $\sigma$-orbital. And the photon numbers in the cavity, $|\alpha_1|^2$ and $|\alpha_2|^2$, correspond to the cavity modes in the $x$ and $y$ directions, respectively. In this context, photons are scattered into both cavities in conjunction with orbital-flip hopping processes, where atoms transition from the $s$-orbital to the $p$-orbital, thus leading to the stabilization of the $p$-orbital superfluid ($\mathrm{SF}_{p_x+p_y}$) and Mott-insulating  ($\mathrm{MI}_{p_x+p_y}$) phases. 

For fixed detuning $\Delta_{1, 2} = 0$, we systematically increase the pump laser strength $V_p$, as illustrated in Fig.~\ref{figure2}. For smaller $V_p$, the photon numbers of both cavities remain zero ($|\alpha_{1, 2}|^2=0$), signifying a normal phase (NP) with only $s$-orbital population. Upon surpassing a critical threshold of the pumping strength $V_p$, photons are symmetrically scattered into all the cavities ($|\alpha_1|^2=|\alpha_2|^2\neq 0$), and the system enters into a new $Z_2\times Z_2$ symmetry-broken $p$-orbital superfluid phase (SF$_{p_x+p_y}$). Notably, the atoms in $s$-orbital are scattered in to $p_x$- and $p_y$-orbitals symmetrically with $n_{p_x}/n_s=n_{p_y}/n_s\neq 0$. As the pumping intensity continues to increase, the enhanced collective scattering effect, coupled with the standing wave potential generated by the cavity fields in both $x$ and $y$ directions,  suppresses the atomic tunneling. This leads to a transition from the $p$-orbital superfluid to $p$-orbital Mott-insulating phases, characterized by the vanishing superfluid order parameters ($|\phi_{p_x, p_y}|=0$) and non-zero photon numbers in both cavities ($|\alpha_1|^2=|\alpha_2|^2\neq 0$). Calculations conducted under the symmetric condition ($\Delta_1=\Delta_2$) indicate that photons are evenly distributed into the two cavities, thereby leading to a degenerate $p_x$- and $p_y$-orbital superradiant phase. This is quite different from our previous work, where atoms can be selectively transferred to the $p_x$-orbital or the $d_{xy}$-orbital states~\cite{PhysRevA.106.023315}. 

\begin{figure}[tbp]
\includegraphics[width=1\linewidth]{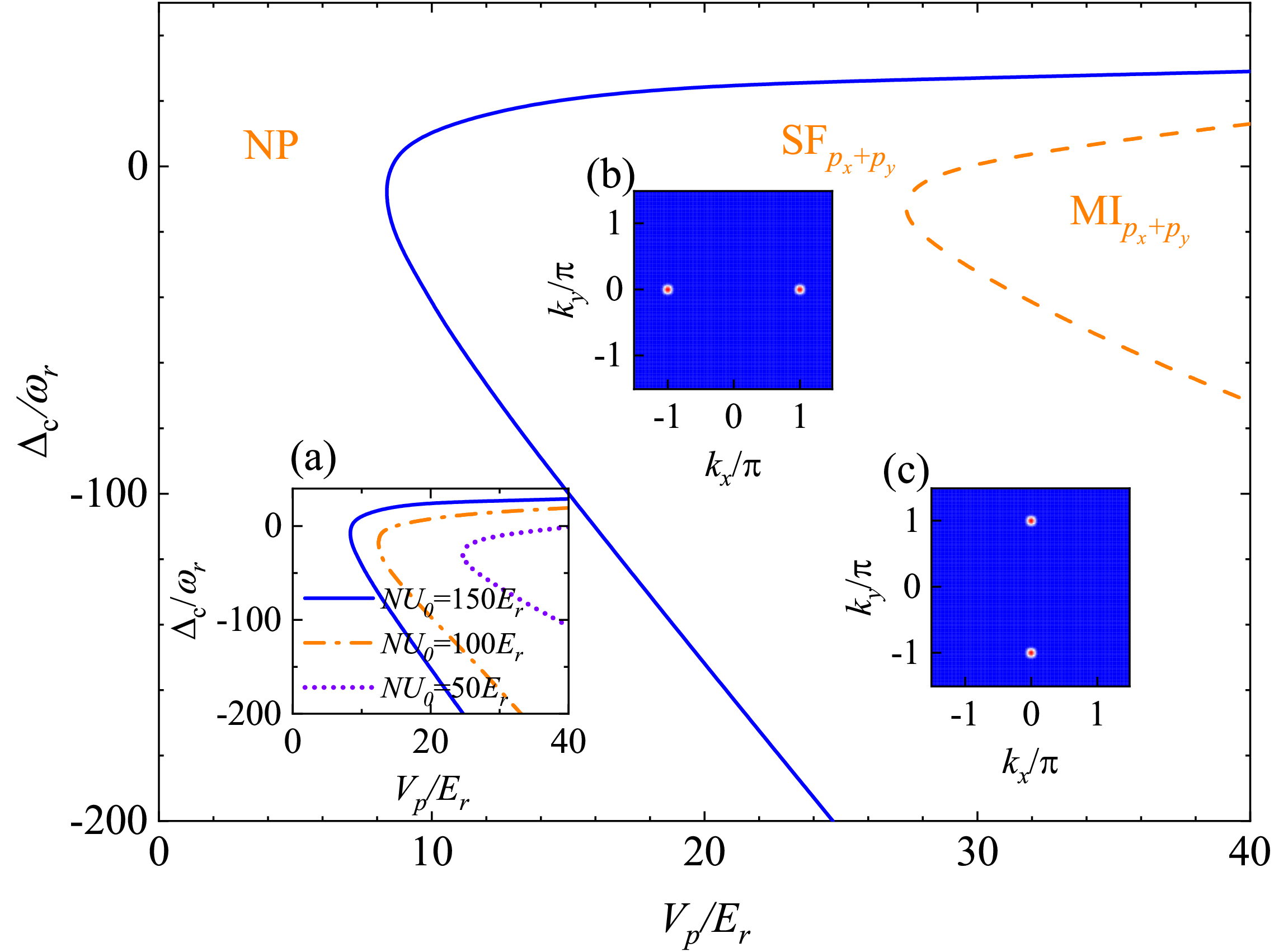}
\caption{Phase diagram of ultarcold bosonic gases coupled to two orthogonal optical cavities, pumped by a blue-detuned running wave laser, obtained from bosonic dynamical mean-field theory. Here, the $p$-orbital superfluid (SF$_{p_x+p_y}$) and Mott-insulating  (MI$_{p_x+p_y}$) phases with superradiance were found. Inset: (a) Phase boundaries of the normal (NP) and $p$-orbital superfluid (SF$_{p_x+p_y}$) phases, where the region of SF$_{p_x+p_y}$ shrinks with the decrease of the atom-cavity coupling. (b)(c) Momentum distributions for the $p_x$- and $p_y$-orbitals, respectively. Other parameters are $N\times U_{0}=150\,E_{r}$ (main figure), and $\protect\kappa =40\,\protect\omega _{r}$. }
\label{figure3}
\end{figure}

We map out the complete phase diagram as a function of the pump strength $V_p$ and cavity detuning $\Delta_{1,2}=\Delta_c$ for fixed $N\times U_0=150E_r$ as shown in Fig.~\ref{figure3}. There exist rich quantum phases, including the NP, SF$_{p_x+p_y}$, and MI$_{p_x+p_y}$ phases. Fig.~\ref{figure3}(a) shows the phase boundaries of NP and SF$_{p_x+p_y}$ for different $N\times U_0$, which indicates that the SF$_{p_x+p_y}$ region shrinks with smaller $N\times U_0$ as the coupling between atoms and cavities is weakened. Momentum distributions for $p_x$- and $p_y$-orbitals in the SF$_{p_x+p_y}$ phase are shown in Figs.~\ref{figure3}(b) and (c), respectively. As expected, the $p_x$-orbital state condensate at $\mathbf{k}=(\pm\pi,0)$, and $\mathbf{k}=(0,\pm\pi)$ for the $p_y$-orbital one, in contrast to the $s$-orbital state condensing at $\mathbf{k}=(0,0)$. Subsequently, the experiment is capable of differentiating atoms across distinct orbital bands utilizing band-mapping techniques~\cite{PhysRevLett.74.1542,PhysRevLett.87.160405,PhysRevLett.94.080403,PhysRevLett.99.200405}.

\subsection{\label{sec:level3B} $p$-orbital self-organization for $\Delta_1\neq\Delta_2$}
\begin{figure*}[tbp]
	\includegraphics[width=1\linewidth]{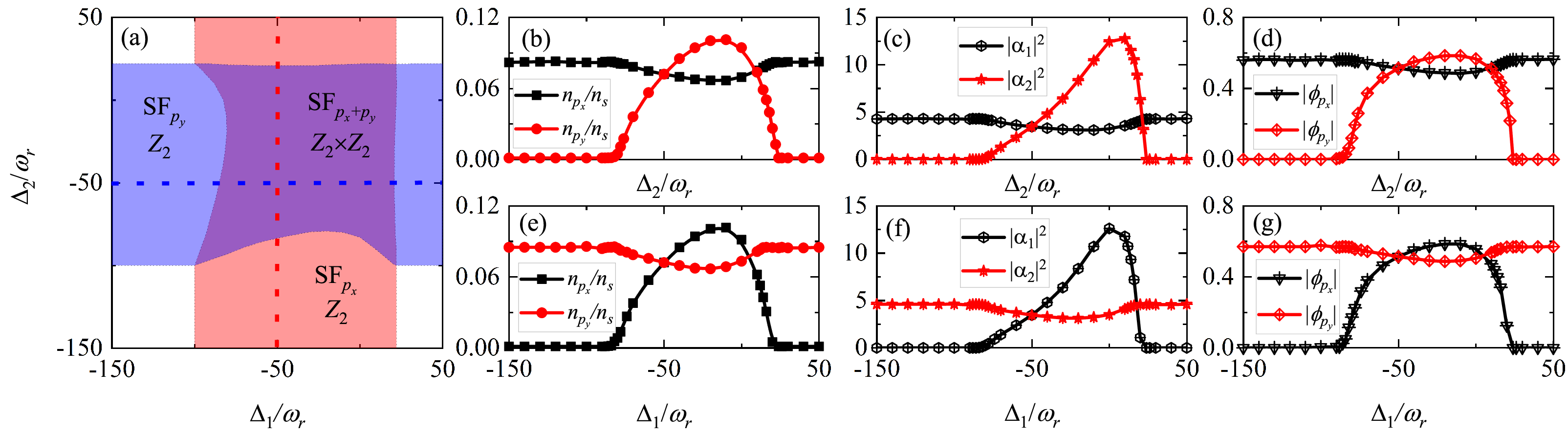}
	\caption{(a) Phase diagram of ultracold bosonic gases trapped in two orthogonal asymmetric optical cavities driven by a blue-detuned laser, revealing various many-body phases with broken different symmetries, including the $p_x$-orbital superfluid ($\mathrm{SF}_{p_x}$), $p_y$-orbital superfluid ($\mathrm{SF}_{p_y}$), and $p$-orbital superfluid ($\mathrm{SF}_{p_x+p_y}$) phases. (b)(e) The ratio of atom population on different orbital states as a function of (b) $\Delta_{2}$, and (e) $\Delta_{1}$. (c)(f) Photon numbers of the cavities in the $x$, and $y$ directions, respectively. (d)(g) Superfluid order parameters for the $p_x$- and $p_y$-orbital states as a function of (d) $\Delta_{2}$,  and (g) $\Delta_{1}$. (b-d) are cut from (a), indicated by the red dotted line for fixed $\Delta_1=-50\omega_{r}$. (e-g) are cut from (a), denoted by the blue dotted line for fixed $\Delta_2=-50\omega_{r}$. Photons scattered into one cavity weaken scattering photons into another one is observed. Other parameters are
	$\protect\kappa =40\,\protect\omega _{r}$, and $N\times U_{0}=150\,E_{r}$.}
	\label{figure4}
\end{figure*}

For the case where $\Delta_1 \neq \Delta_2$, photons are asymmetrically scattered into cavities along the $x$ and $y$ directions, which leads to the stabilization of additional quantum phases that break distinct symmetries, as illustrated in Fig.~\ref{figure4}(a).

There are four distinct quantum phases, including the normal (NP), $p_x$-orbital superfluid (SF$_{p_x}$), $p_y$-orbital superfluid (SF$_{p_y}$), and $p$-orbital superfluid (SF$_{p_x+p_y}$) phases. We construct the complete phase diagram as a function of cavity detunings $\Delta_1$ and $\Delta_2$, with fixing the pump strength at $V_p=15E_r$ and $N\times U_0=150E_r$. For smaller or larger values of $\Delta_1$ (light blue region), photons are preferentially scattered into the cavity in the $y$ direction, which favors the SF$_{p_y}$ phase with broken $Z_2$ symmetry. Similarly, for smaller or larger values of $\Delta_2$ (light red region), photons are confined to the $x$-directional cavity, and the system remains in the SF$_{p_x}$ phase with broken $Z_2$ symmetry. In the intermediate region (overlap area), photons are scattered into both cavities, enabling orbital-flipping processes that transfer population from the $s$-orbital to both the $p_x$- and $p_y$-orbitals. This stabilizes the SF$_{p_x+p_y}$ phase with broken $Z_2\times Z_2$ symmetry.

Here, we employ various order parameters to characterize the phase transitions, including the orbital population ratios $n_{p_x, p_y}/n_{s}$ [Figs.~\ref{figure4}(b)(e)], the photon numbers $|\alpha_{1,2}|^2$ [Figs.~\ref{figure4}(c)(f)], and the superfluid order parameters $|\phi_{p_x, p_y}|$ [Figs.~\ref{figure4}(d)(g)]. The order parameters as a function of $\Delta_2$ in Figs.~\ref{figure4}(b-d) are extracted from Fig.~\ref{figure4}(a) (red dotted line) for fixed $\Delta_1=-50\omega_{r}$. Initially, the system resides in the $\mathrm{SF}_{p_x}$ phase. As $\Delta_2$ increases, the number of photons scattered into the $y$-directional cavity mode initially increases and subsequently decreases back to zero. The buildup of the $p_y$-orbital phase suppresses the scattering of photons into the $x$-directional cavity mode, accompanied by the reduction of the occupation on the $p_x$-orbital and the superfluid order parameter ($|\phi_{p_x}|$). This observation indicates a competitive relationship between the $\mathrm{SF}_{p_x}$ and $\mathrm{SF}_{p_y}$ phases, which is analogous to the competition of the two orders in Ref.~\cite{morales_coupling_2018}. 
Further confirmation comes from Figs.~\ref{figure4}(e–g), where order parameters are shown as a function of $\Delta_1$ for fixed $\Delta_2=-50\omega_{r}$ [blue dotted line in Fig.~\ref{figure4}(a)]. It exhibits mutual  suppression of photon scattering into distinct cavity modes and competitive population redistribution in the high-orbital states.

The competition between different phases is also reflected in Fig.~\ref{figure4}(a), the upper and right boundaries of the overlapping region remain unaffected, attributed to the rapid increase in photon scattering into the cavity modes. Conversely, the left and bottom boundaries exhibit a marked suppression, as the increase in photon scattering into the cavity modes in these regions is relatively slow, as shown in Figs.~\ref{figure4}(c) and (f).

\subsection{\label{sec:level3C} Orbital-density wave in superradiant phase}
Unlike the charge-density-wave superradiant phases emerging under a red-detuned pumping, a blue-detuned cavity induces self-organized orbital-density waves in the excited atoms. This occurs via cavity-mediated orbital-flipping processes, as shown in Fig.~\ref{figure1}, ensuring a uniform local filling for the $s$-, $p_x$-, and $p_y$-orbital states, respectively.

Here, we treat the orbital degrees of freedom as pseudospins, for instance, any two of the three $s$-, $p_x$-, and $p_y$-orbital states essentially achieve a pseudospin-1/2 system in the cavity-induced optical lattices. The local magnetism is defined by
${\bf \hat{S}}^{\sigma_1\sigma_2}_i = \hat{b}^\dagger_{i,\sigma} {\bf F}_{\sigma\sigma'} \hat{b}_{i,\sigma'} $,
where ${\bf F}_{\sigma\sigma^\prime}$ is the matrix for spin-1/2 particles, {\it i.e.} $\hat{S}^{\sigma\sigma'}_{i,x}= 1/2  ({\hat{b}_{i,\sigma}}^\dagger {\hat{b}_{i,\sigma'}} + {\hat{b}_{i,\sigma'}}^\dagger {\hat{b}_{i,\sigma}}) $, $\hat{S}^{\sigma\sigma'}_{i,y}=i/2 (-{ \hat{b}_{i,\sigma}}^\dagger { \hat{b}_{i,\sigma'}} + {\hat{b}_{i,\sigma'}}^\dagger {\hat{b}_{i,\sigma}}) $, and $\hat{S}^{\sigma\sigma'}_{i,z}= 1/2  ({\hat{b}_{i,\sigma}}^\dagger {\hat{b}_{i,\sigma}} - {\hat{b}_{i,\sigma'}}^\dagger  {\hat{b}_{i,\sigma'}} )$.

\begin{figure}[tbp]
\includegraphics[width=1.0\linewidth]{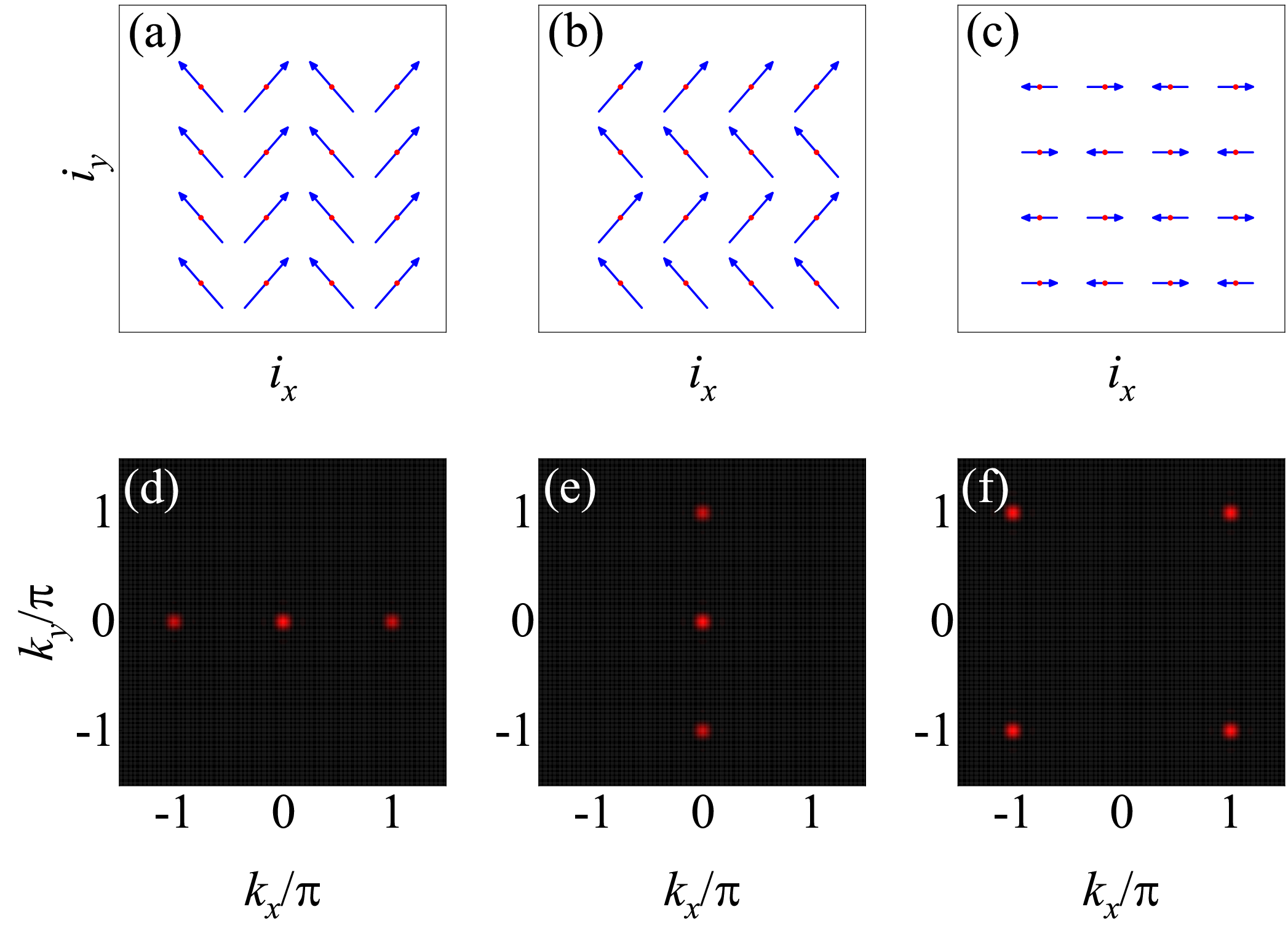}
	\caption{\label{figure5} (a-c) Real-space orbital-order distributions in the $\mathrm{SF}_{p_x+p_y}$ phase for (a) $(\langle\hat{S}^{s p_x}_{i,x}\rangle$, $\langle\hat{S}^{s p_x}_{i,z}\rangle)$, (b) $(\langle\hat{S}^{s p_y}_{i,x}\rangle$, $\langle\hat{S}^{s p_y}_{i,z}\rangle)$, and (c) $(\langle\hat{S}^{p_x p_y}_{i,x}\rangle$, $\langle\hat{S}^{p_x p_y}_{i,z}\rangle)$, obtained by bosonic dynamical mean-field theory. (d-f) Static spin structure factor in the $\mathrm{SF}_{p_x+p_y}$ phase for (d) $S^{s p_x}_{\bf k}$, (e) $S^{s p_y}_{\bf k}$, and (f) $S^{p_x p_y}_{\bf k}$, respectively. The parameters are $N\times U_0 = 150\,E_r$, $\Delta_{1,2}=0\,\omega_r$, and $V_p=25E_r$.}
\end{figure}
The real-space orbital-order distributions $(S^{\sigma_1\sigma_2}_{i,x}$,
$S^{\sigma_1\sigma_2}_{i,z})$ in the $\mathrm{SF}_{p_x+p_y}$ phase are shown in the top row of Fig.~\ref{figure5}. We find that the orbital-density waves behave as
\begin{eqnarray}
    \langle\hat{S}^{sp_x}_{j,x}\rangle&=&(-1)^{j_x+i_x}\langle\hat{S}^{sp_x}_{i,x}\rangle,\\
    \langle\hat{S}^{sp_y}_{j,x}\rangle&=&(-1)^{j_y+i_y}\langle\hat{S}^{sp_y}_{i,x}\rangle,\\
    \langle\hat{S}^{p_xp_y}_{j,x}\rangle&=&(-1)^{j_x+i_x+j_y+i_y}\langle\hat{S}^{p_xp_y}_{i,x}\rangle.
\end{eqnarray} 
On the other hand, we observed that $\langle\hat{S}^{sp_x}_{j,x}\rangle$ ($\langle\hat{S}^{sp_y}_{j,x}\rangle$) remains uniform along the $y\ (x)$ direction, whereas along the $x\ (y)$ direction, the values of $\langle\hat{S}^{sp_x}_{j,x}\rangle$ ($\langle\hat{S}^{sp_y}_{j,x}\rangle$) at adjacent sites are opposite in sign. For $\langle\hat{S}^{p_xp_y}_{j,x}\rangle$, its value alternates in sign between adjacent lattice sites along both the $x$ and the $y$ directions. These different orbital-density waves are induced by the on-site intra-orbital scattering processes, which can be understood by deriving the two-site model as in our previous work~\cite{PhysRevA.106.023315}. Here, we derive the cavity-mediated spin-spin interactions (see Appendix B), and obtain an effective model
\begin{eqnarray}
\hat{H}^{sp_x}_\mathrm{spin}&\propto\sum_{ii'}\chi_{sp_x}(-1)^{i_x+i'_x} \hat{S} ^{sp_x}_{i,x}\hat{S} ^{sp_x}_{i^\prime,x},
\end{eqnarray}
\begin{align}
\hat{H}^{sp_y}_\mathrm{spin}&\propto\sum_{ii'} \chi_{sp_y}(-1)^{i_y+i'_y} \hat{S} ^{sp_y}_{i,x}\hat{S} ^{sp_y}_{i^\prime,x},\\
\hat{H}^{p_xp_y}_\mathrm{spin}&\propto\sum_{i}\chi_{p_xp_ysp_xsp_y}(-1)^{i_x+i_y} \hat{S} ^{p_xp_y}_{i,x}\notag\\
&+ \sum_{ii'}(\chi_{p_xp_ysp_x}+\chi_{p_xp_ysp_y})\notag\\
&\times(-1)^{i_x+i_y + i'_x+i'_y} \hat{S} ^{p_xp_y}_{i,x}
\hat{S} ^{p_xp_y}_{i^{\prime},x}.
\end{align}
Obviously, the spin-spin interactions are mediated by factors $(-1)^{i_x+i'_x}$, $(-1)^{i_y+i'_y}$, $(-1)^{i_x+i_y}$, and $(-1)^{i_x+i_y + i'_x+i'_y}$, which are associated with the lattice sites. These factors play a crucial role in stabilizing different orbital-density waves, thereby lowering the system's energy. The spin structure factor is defined as $ S^{^{\sigma_1\sigma_2}}_{\bf k} =\frac{1}{N_\mathrm{lat}}\left| \sum_i \langle {\bf \hat{S}}^{\sigma_1\sigma_2}_i\rangle e^{i  {\bf k} \cdot  {\bf r}_i} \right| $ \cite{PhysRevLett.109.085302} and shown in the bottom row of Fig.~\ref{figure5}.

\section{\label{sec:level4} Conclusions}
We investigate strongly correlated ultracold bosonic gases that are coupled to two orthogonal cavities and pumped by a blue-detuned running wave laser. Our findings reveal that atoms initially in the $s$-orbital state can be scattered into $p_x$- and $p_y$-orbital states, exhibiting symmetric or asymmetric behavior based on the frequencies of the cavities. For the symmetric configuration, we observe that atoms are evenly scattered into the $p_x$- and $p_y$-orbitals. Notably, for the asymmetric case, we find that photons scattered into one cavity mode can weaken the scattering of photons into the orthogonal mode. In addition, we observe that the emergence of high-orbital self-organization is accompanied by orbital-density waves that break distinct symmetries. 
Future research could explore the coupling of three cavity modes, where orbital frustrations are at play within the triangular lattice induced by the cavities, potentially leading to a more diverse array of multi-orbital phenomena.

\section{\label{sec:level7} Acknowledgements} This work is supported by the National Natural Science Foundation of China under ation of China under Grants Nos. 12374252, 12074431, and 12174130, and
the Science and Technology Innovation Program of Hunan
Province under Grant No. 2024RC1046. We acknowledge the ChinaHPC for providing HPC resources that have contributed to the research results
reported within this paper.

\begin{widetext}
	\begin{center}
		 \section{\label{sec:level8} Appendix}
	\end{center}
	\renewcommand{\theequation}{A\arabic{equation}}
	\renewcommand{\thesection}{S-\arabic{section}}
	\renewcommand{\bibnumfmt}[1]{[S#1]}
	\renewcommand{\citenumfont}[1]{S#1}
	\setcounter{equation}{0}

\subsection{\label{sec:level8A} Extended Bose-Hubbard model}

The two orthogonal cavities in $x$ and $y$ directions are pumped by a blue-detuned running wave laser in the $z$ direction
\begin{eqnarray}
	E(z) = E_0\mathrm{cos}(k_pz-w_pt),
\end{eqnarray}
where $k_p$ is the wave vector of the pump laser. Then the effective interaction for the electric field and atom coupled system is
\begin{eqnarray}
	\hat{H}&=&\frac{1}{2}\hbar\omega_a\hat{\sigma}_z + \hbar\omega_1 \hat{a}^\dagger_1 \hat{a}_1 + \hbar\omega_2 \hat{a}^\dagger_2 \hat{a}_2 + \frac{\hbar\Omega_p}{2}(\hat{\sigma}^+ + \hat{\sigma}^-)\left[e^{i(k_pz-\omega_p t)}+e^{-i(k_pz-\omega_p t)} \right]
	\nonumber\\
	&+& \hbar\lambda_1(\hat{\sigma}^+\hat{a}_1 + \hat{\sigma}^-\hat{a}^\dagger_1) + \hbar\lambda_2(\hat{\sigma}^+\hat{a}_2 + \hat{\sigma}^-\hat{a}^\dagger_2),
\end{eqnarray}
where $\lambda_{1}=g_0\cos(k_c x)$, $\lambda_{2}=g_0\cos(k_c y)$, $\hat{\sigma}^{+}=|e\rangle\langle g|$, and $\hat{\sigma}^{-}=|g\rangle\langle e|$ with $|g\rangle$ and $|e\rangle$ being the ground and excited states, respectively. By transforming into the reference frame that rotates at the frequency $w_p$, one yields
\begin{eqnarray}
	\hat{H}&=&\frac{\bm{p}^2}{2m} -\hbar\Delta_a\hat{\sigma}^+ \hat{\sigma}^- -\hbar\Delta_1 \hat{a}^\dagger_1 \hat{a}_1 + \hbar\Delta_2 \hat{a}^\dagger_2 \hat{a}_2 + \frac{\hbar\Omega_p}{2} \left( \hat{\sigma}^+ e^{i k_pz}+ \hat{\sigma}^- e^{-i k_pz}\right)\nonumber\\
	&+& \hbar\lambda_1(\hat{\sigma}^+\hat{a}_1 + \hat{\sigma}^-\hat{a}^\dagger_1) + \hbar\lambda_2(\hat{\sigma}^+\hat{a}_2 + \hat{\sigma}^-\hat{a}^\dagger_2).
\end{eqnarray}
Based on the single-particle Hamiltonian, we construct the many-body counterpart expressed in second-quantized form
\begin{eqnarray}\label{H_new_1}
	\hat{H}&=&\int d{\bf r} \left[ \hat{\Phi}^\dagger_g ({\bf r})\left(-\frac{\hbar^2\nabla^2}{2m} \right) \hat{\Phi}_g({\bf r}) + \hat{\Phi}^\dagger_e ({\bf r})\left(-\frac{\hbar^2\nabla^2}{2m} - \hbar\Delta_a\right) \hat{\Phi}_e({\bf r})
	\right]\nonumber\\
	&+&\int d{\bf r} \left[ \hat{\Phi}^\dagger_e ({\bf r})\left(\frac{\hbar\Omega_p}{2}e^{i k_pz} + \hbar\lambda_1 \hat{a}_1 + \hbar\lambda_2 \hat{a}_2 \right) \hat{\Phi}_g({\bf r})+\mathrm{H.c.} \right] - \hbar\Delta_1\hat{a}^\dagger_1 \hat{a}_1 - \hbar\Delta_2\hat{a}^\dagger_2 \hat{a}_2,
\end{eqnarray}
where $\Phi_g$ and $\Phi_e$ are 
the field operators for the ground and excited states, respectively.
Adiabatically eliminating the excited state of the atom for large detuning $\Delta_{1,2}$ and taking the interaction between atoms into account, we finally obtain
\begin{eqnarray}
	\hat{H}&=&\int d{\bf r} \left[ \hat{\Phi}^\dagger_g ({\bf r})\left(-\frac{\hbar^2\nabla^2}{2m}  + V({\bf r}) \right) \hat{\Phi}_g({\bf r})\right]  - \hbar\Delta_1\hat{a}^\dagger_1 \hat{a}_1 - \hbar\Delta_2\hat{a}^\dagger_2 \hat{a}_2 \nonumber\\
	&+& \int d{\bf r} \left[\hat{\Phi}^\dagger_g ({\bf r}) \left(\frac{\sqrt{V_pU_0}}{2}\cos(kx)\left(\hat{a}_1+\hat{a}^\dagger_1 \right) + \frac{\sqrt{V_pU_0}}{2}\cos(ky)\left(\hat{a}_2 +\hat{a}^\dagger_2  \right)  \right)  \hat{\Phi}_g({\bf r}) \right]\nonumber\\
	&+& \int d{\bf r} \left[\hat{\Phi}^\dagger_g ({\bf r}) U_0\cos(kx)\cos(ky)\left(\hat{a}^\dagger_1 \hat{a}_2+\hat{a}^\dagger_2 \hat{a}_1 \right)   \hat{\Phi}_g({\bf r}) \right]\nonumber\\
	&+&\frac{g}{2}\int d{\bf r} \hat{\Phi}^\dagger_g({\bf r})\hat{\Phi}^\dagger_g({\bf r})\hat{\Phi}_g({\bf r})\hat{\Phi}_g({\bf r}),
\end{eqnarray}
where $V_p=\hbar \Omega^2_p/\Delta_a$, $U_0=\hbar g^2_0/\Delta_a $,  and $k_c\approx k_p=k$. Note here that the Hamiltonian is restricted to two-dimensional case in the $xy$ plane by applying a strong standing wave field in the $z$ direction with $V_z=50E_r$. As the two cavities are pumped by a blue-detuned pump laser, atoms in the cavity field can be treated as a low-field seeker, resulting in an atom position shift of $\lambda/4$ in both the $x$ and $y$ directions compared to the red-detuned scenario. By defining the atom's position as the origin of the coordinate system, the effective Hamiltonian can be expressed as
\begin{eqnarray}
	\hat{H}&=&\int d{\bf r} \left[ \hat{\Phi}^\dagger ({\bf r})\left(-\frac{\hbar^2\nabla^2}{2m}  + V({\bf r}) \right) \hat{\Phi}({\bf r})\right]  - \hbar\Delta_1\hat{a}^\dagger_1 \hat{a}_1 - \hbar\Delta_2\hat{a}^\dagger_2 \hat{a}_2 \nonumber\\
	&+& \int d{\bf r} \left[\hat{\Phi}^\dagger ({\bf r}) \left(\frac{\sqrt{V_pU_0}}{2}\sin(kx)\left(\hat{a}_1+\hat{a}^\dagger_1 \right) + \frac{\sqrt{V_pU_0}}{2}\sin(ky)\left(\hat{a}_2 +\hat{a}^\dagger_2  \right)  \right)  \hat{\Phi}({\bf r}) \right]\nonumber\\
	&+& \int d{\bf r} \left[\hat{\Phi}^\dagger ({\bf r}) U_0\sin(kx)\sin(ky)\left(\hat{a}^\dagger_1 \hat{a}_2+\hat{a}^\dagger_2 \hat{a}_1 \right)   \hat{\Phi}({\bf r}) \right]\nonumber\\
	&+&\frac{g}{2}\int d{\bf r} \hat{\Phi}^\dagger({\bf r})\hat{\Phi}^\dagger({\bf r})\hat{\Phi}({\bf r})\hat{\Phi}({\bf r}),
\end{eqnarray}
where we replace $\hat{\Phi}_g({\bf r})$ by $\hat{\Psi}({\bf r})$ for simplify.

In sufficiently deep optical lattices, the atomic field operator can be expanded on the Wannier basis $\hat{\Psi}({\bf r})= \sum_{i,\sigma} \hat{b}_{i,\sigma}w_{\sigma}({\bf r}-{\bf r}_i)$, where operator $\hat{b}_{i,\sigma}$ ($\hat{b}^\dagger_{i,\sigma}$) annihilate (create) a Wannier state $\sigma$ at site $i$, and the Wannier function $w_{\sigma}({\bf r}-{\bf r}_i)$ centered at ${\bf r} = {\bf r}_i$ for the $s$-, $p_x$- and $p_y$-orbital states, respectively. Then, the many-body system can be effectively described by an extended Bose-Hubbard model
\begin{eqnarray}
\hat{H} &=&  - \sum_{\langle i, j\rangle,\sigma}J^{ij}_{
	\sigma\sigma}\left(\hat{b}^\dagger_{i,\sigma} \hat{b}_{j,\sigma}+ {\rm H.c.}\right) +\sum\limits_{i,\sigma _{1}\sigma _{2}\sigma _{3}\sigma _{4}}\frac{%
	U_{\sigma _{1}\sigma _{2}\sigma _{3}\sigma _{4}}}{2}\hat{b}_{i,\sigma
	_{1}}^{\dag }\hat{b}_{i,\sigma _{2}}^{\dag }\hat{b}_{i,\sigma _{3}}\hat{b}%
	_{i,\sigma _{4}} -\sum_{i,\sigma}\mu_{\sigma}\hat{b}^\dagger_{i,\sigma} \hat{b}_{i,\sigma}\nonumber\\
	&+&2\mathrm{Re}[\alpha_1]\sum_{ij} (-1)^{i_x}\left(J^{ij}_{sp_x} \hat{b}^\dagger_{i,s} \hat{b}_{j,p_x} + \mathrm{H.c.} \right)+2\mathrm{Re}[\alpha_2]\sum_{ij} (-1)^{i_y}\left(J^{ij}_{sp_y} \hat{b}^\dagger_{i,s} \hat{b}_{j,p_y} + \mathrm{H.c.} \right)\nonumber\\
	&+&2\left( \mathrm{Re}[\alpha_1]*\mathrm{Re}[\alpha_2] + \mathrm{Im}[\alpha_1]*\mathrm{Im}[\alpha_2] \right)\sum_{ij} (-1)^{i_x + i_y}\left(J^{ij}_{p_xp_y} \hat{b}^\dagger_{i,p_x} \hat{b}_{j,p_y} + \mathrm{H.c.} \right)-\hbar |\alpha_1|^2 \Delta_1 - \hbar |\alpha_2|^2 \Delta_2, 
\end{eqnarray}
where $\mu_\sigma \equiv J^{ii}_{\sigma\sigma}$ is the chemical potential,  $\langle i,j\rangle$ represents the nearest
neighbor sites $i,j$. The cavity field can be simplified by the coherent-state approximation $\alpha_1 =\langle \hat{a}_1\rangle$ and $\alpha_2 =\langle \hat{a}_2\rangle$, while the cavity modes are macroscopically populated~\cite{PhysRevLett.107.140402}. We then determine the cavity fields in the steady state for a large decay rate $\kappa$,
\begin{eqnarray}
	\alpha_1 =\frac{\sum_{i}\left\langle (-1)^{i_x}J^{ii}_{sp_x}%
	\hat{b}_{i,s}^{\dagger }\hat{b}_{i,p_x}+ \mathrm{H.c.} +(-1)^{i_x+i_y}\alpha_2 J^{ii}_{p_{x}p_{y}}\left(\hat{b}%
	_{i,p_x}^{\dagger }\hat{b}_{i,p_{y}} +\mathrm{H.c.}\right)\right\rangle} { \Delta _{1}-\sum_{i,\sigma
	}\langle J_{1,\sigma }\hat{b}_{i,\sigma }^{\dagger }\hat{b}_{i,\sigma }\rangle
	+i\kappa },
\end{eqnarray}
\begin{eqnarray}
	\alpha_2 =\frac{\sum_{i}\left\langle (-1)^{i_y}J^{ii}_{sp_{y}}\hat{b}%
		_{i,s}^{\dagger }\hat{b}_{i,p_{y}}+\mathrm{H.c.} +(-1)^{i_x+i_y}\alpha_1 J^{ii}_{p_{x}p_{y}}\left(\hat{b}%
		_{i,p_x}^{\dagger }\hat{b}_{i,p_{y}} +\mathrm{H.c.}\right)\right\rangle} { \Delta _{2}-\sum_{i,\sigma
		}\langle J_{2,\sigma }\hat{b}_{i,\sigma }^{\dagger }\hat{b}_{i,\sigma }\rangle
		+i\kappa },
\end{eqnarray}
 with $J_{1,\sigma} =\int d \mathbf{r} w^\ast_{\sigma}\left(\mathbf{r}-\mathbf{r}_{i}\right)U_{\rm 0}\, {\sin^2(kx)} w_{\sigma}\left(\mathbf{r}-\mathbf{r}_{i}\right)$, $J_{2,\sigma} =\int d \mathbf{r} w^\ast_{\sigma}\left(\mathbf{r}-\mathbf{r}_{i}\right)U_{\rm 0}\, {\sin^2(ky)} w_{\sigma}\left(\mathbf{r}-\mathbf{r}_{i}\right)$.
Other coupling matrix elements are
\begin{equation}
	J^{ij}_{\sigma\sigma} =-\int d \mathbf{r} w^\ast_{\sigma}\left(\mathbf{r}-\mathbf{r}_{i}\right)\left(-\frac{\hbar^2 \nabla^{2}}{2 m} + V_{\rm lat}\right) w_{\sigma}\left(\mathbf{r}-\mathbf{r}_{j}\right),
\end{equation}
\begin{equation}
	J^{ij}_{p_xp_y}=\int d \mathbf{r}w^\ast_{p_y}\left(\mathbf{r}-\mathbf{r}_{i}\right) U_0  \sin (k x)\sin (k_c y) w_{p_x}\left(\mathbf{r}-\mathbf{r}_{j}\right),
\end{equation}
\begin{equation}
	J^{ij}_{sp_x}=\int d \mathbf{r}w^\ast_{p_x}\left(\mathbf{r}-\mathbf{r}_{i}\right) \frac{\sqrt{V_p U_0}}{2}   \sin (k x) w_{s}\left(\mathbf{r}-\mathbf{r}_{j}\right),
\end{equation}
\begin{equation}
	J^{ij}_{sp_y}=\int d \mathbf{r}w^\ast_{p_y}\left(\mathbf{r}-\mathbf{r}_{i}\right) \frac{\sqrt{V_p U_0}}{2}   \sin (k y) w_{s}\left(\mathbf{r}-\mathbf{r}_{j}\right),
\end{equation}
\begin{equation}
	U_{\sigma _{1}\sigma _{2}\sigma _{3}\sigma _{4}}=\int d \mathbf{r} w^\ast_{\sigma_1}\left(\mathbf{r}-\mathbf{r}_{i}\right) w^\ast_{\sigma_2}\left(\mathbf{r}-\mathbf{r}_{i}\right) \frac{4\pi \hbar^2 a_s}{m} w_{\sigma_3}\left(\mathbf{r}-\mathbf{r}_{i}\right) w_{\sigma_4}\left(\mathbf{r}-\mathbf{r}_{i}\right).
\end{equation}
Here, $V_{\rm lat}=(V_{x}+U_0|\alpha_1|^2)\,  {\rm sin}^2(kx)$ in the $x$ direction and $V_{\rm lat}=(V_{y}+U_0|\alpha_2|^2\,  {\rm sin}^2(ky)$ in the $y$ direction, with $V_{x,y}=5E_r $ being an external optical lattice added in the cavity directions to validate the tight-binding model. The onsite interaction terms considered in our work read
\begin{eqnarray}\label{onsite_interaction}
	&&\sum\limits_{i,\sigma _{1}\sigma _{2}\sigma _{3}\sigma _{4}}\frac{%
		U_{\sigma _{1}\sigma _{2}\sigma _{3}\sigma _{4}}}{2}\hat{b}_{i,\sigma
		_{1}}^{\dag }\hat{b}_{i,\sigma _{2}}^{\dag }\hat{b}_{i,\sigma _{3}}\hat{b}%
	_{i,\sigma _{4}}\nonumber\\
	&=&\sum_{i}\bigg[\sum_{\sigma=s,p_x,p_y}\frac{U_{i,\sigma}}{2}\hat{n}_{i,\sigma}(\hat{n}_{i,\sigma}-1) + 2U_{i,sp_x}n_{i,s}n_{i,p_x}+ 2U_{i,sp_y}n_{i,s}n_{i,p_y} + 2U_{i,p_xp_y}n_{i,p_x}n_{i,p_y}\\
	&+&\frac{U_{i,sp_x}}{2}\left(b^\dagger_{i,s}b^\dagger_{i,s}b_{i,p_x}b_{i,p_x} + \rm{H.c.}  \right) + \frac{U_{i,sp_y}}{2}\left(b^\dagger_{i,s}b^\dagger_{i,s}b_{i,p_y}b_{i,p_y} + \rm{H.c.}  \right) + \frac{U_{i,p_xp_y}}{2}\left(b^\dagger_{i,p_x}b^\dagger_{i,p_x}b_{i,p_y}b_{i,p_y} + \rm{H.c.}  \right)
	\bigg].\nonumber
\end{eqnarray}

\subsection{\label{sec:level8b} Derivation of spin-spin interaction mediated by cavity}
Deep in the superradiant phase, the nearest neighbour hopping and scattering can be neglected, and we only consider the onsite terms
\begin{eqnarray}\label{onsite_ham}
	\hat{H} &=&\sum\limits_{i,\sigma _{1}\sigma _{2}\sigma _{3}\sigma _{4}}\frac{%
		U_{\sigma _{1}\sigma _{2}\sigma _{3}\sigma _{4}}}{2}\hat{b}_{i,\sigma
		_{1}}^{\dag }\hat{b}_{i,\sigma _{2}}^{\dag }\hat{b}_{i,\sigma _{3}}\hat{b}%
	_{i,\sigma _{4}} -\sum_{i,\sigma}\mu_{\sigma}\hat{b}^\dagger_{i,\sigma} \hat{b}_{i,\sigma}\nonumber\\
	&+&(\hat{a}^\dagger_1 + \hat{a}_1)\sum_{i} (-1)^{i_x}\left(J^{ii}_{sp_x} \hat{b}^\dagger_{i,s} \hat{b}_{i,p_x} + \mathrm{H.c.} \right)+(\hat{a}^\dagger_2 + \hat{a}_2)\sum_{i} (-1)^{i_y}\left(J^{ii}_{sp_y} \hat{b}^\dagger_{i,s} \hat{b}_{,p_y} + \mathrm{H.c.} \right)\nonumber\\
	&+&(\hat{a}^\dagger_1\hat{a}_2 +\hat{a}^\dagger_2\hat{a}_1)\sum_{i} (-1)^{i_x + i_y}\left(J^{i}_{p_xp_y} \hat{b}^\dagger_{i,p_x} \hat{b}_{i,p_y} + \mathrm{H.c.} \right)-\hbar \hat{a}^\dagger_1\hat{a}_1 \Delta_1 - \hbar \hat{a}^\dagger_2\hat{a}_2\Delta_2, 
\end{eqnarray}
with the cavity field
\begin{eqnarray}
	\hat{a}_1 &=&\frac{\sum_{i} (-1)^{i_x}J^{ii}_{sp_x}%
		\left(\hat{b}_{i,s}^{\dagger }\hat{b}_{i,p_x}+ \mathrm{H.c.} \right)+(-1)^{i_x+i_y}\hat{a}_2 J^{ii}_{p_{x}p_{y}}\left(\hat{b}%
		_{i,p_x}^{\dagger }\hat{b}_{i,p_{y}} +\mathrm{H.c.}\right)} { \Delta _{1}-\sum_{i,\sigma
		} J_{1,\sigma }\hat{b}_{i,\sigma }^{\dagger }\hat{b}_{i,\sigma }
		+i\kappa }\nonumber\\
	&=&\frac{\sum_{i}(-1)^{i_x}J^{ii}_{sp_x}S_{i,x}^{sp_x} +(-1)^{i_x+i_y}\hat{a}_2 J^{ii}_{p_{x}p_{y}}S_{i,x}^{p_xp_y}} { 0.5(\Delta _{1}-\sum_{i,\sigma
		} J_{1,\sigma }\hat{b}_{i,\sigma }^{\dagger }\hat{b}_{i,\sigma }
		+i\kappa )},
\end{eqnarray}
\begin{eqnarray}
	\hat{a}_2 &=&\frac{\sum_{i} (-1)^{i_y}J^{ii}_{sp_{y}}\left(\hat{b}%
		_{i,s}^{\dagger }\hat{b}_{i,p_{y}}+\mathrm{H.c.}\right) +(-1)^{i_x+i_y}\hat{a}_1 J^{ii}_{p_{x}p_{y}}\left(\hat{b}%
		_{i,p_x}^{\dagger }\hat{b}_{i,p_{y}} +\mathrm{H.c.}\right)} { \Delta _{1}-\sum_{i,\sigma
		} J_{2,\sigma }\hat{b}_{i,\sigma }^{\dagger }\hat{b}_{i,\sigma }
		+i\kappa }\nonumber\\
	&=&\frac{\sum_{i}(-1)^{i_y}J^{ii}_{sp_y}S_{i,x}^{sp_y} +(-1)^{i_x+i_y}\hat{a}_1 J^{ii}_{p_{x}p_{y}}S_{i,x}^{p_xp_y}} { 0.5(\Delta _{1}-\sum_{i,\sigma} J_{2,\sigma }\hat{b}_{i,\sigma }^{\dagger }\hat{b}_{i,\sigma }+i\kappa )}.
\end{eqnarray}
Here, $\hat{a}_1$ and $\hat{a}_2$ are solved by
\begin{eqnarray}
	\hat{a}_1 =(2 \tilde{\Delta}_{2}\sum_{i} (-1)^{i_x}J^{ii}_{sp_x} S ^{sp_x}_{i,x}  + { 4\sum_{i,i^\prime}(-1)^{i_x+i_y+i^\prime_y} J^{i i}_{p_{x}p_{y}}J^{i^\prime i^\prime}_{sp_{y}} S ^{p_xp_y}_{i,x} 	S ^{sp_y}_{i^\prime,x}   } )/\mathcal{Z},
\end{eqnarray}
\begin{eqnarray}
	\hat{a}_2 =( 2 \tilde{\Delta}_{1} \sum_{i} (-1)^{i_y}J^{ii}_{sp_y} S ^{sp_y}_{i,x}  +  4\sum_{i,i^\prime}(-1)^{i_x+i_y+i^\prime_x} J^{i i}_{p_{x}p_{y}}J^{i^\prime i^\prime}_{sp_{x}} S ^{p_xp_y}_{i,x} 	S ^{sp_x}_{i^\prime,x}   )/\mathcal{Z},
\end{eqnarray}
where $\tilde{\Delta}_{1(2)}=\Delta _{1(2)}-\sum_{i,\sigma}\langle J_{1(2),\sigma }\hat{b}_{i,\sigma }^{\dagger }\hat{b}_{i,\sigma }\rangle
+i\kappa$, and $\mathcal{Z}= \tilde{\Delta}_{1} \tilde{\Delta}_{2}-4\sum_{i,i^\prime}(-1)^{i_x+i_y+i^\prime_x+i^\prime_y} J^{i i}_{p_{x}p_{y}}J^{i^\prime i^\prime}_{p_{x}p_{y}}S ^{p_xp_y}_{i,x}S ^{p_xp_y}_{i^\prime,x}$. Substituting into Eq.~(\ref{onsite_ham}), we obtain an effective atom-only Hamiltonian
\begin{eqnarray}
	\hat{H} &=&\sum_{ii'}\chi_{sp_x}(-1)^{i_x+i'_x} S ^{sp_x}_{i,x}S ^{sp_x}_{i^\prime,x} +\sum_{ii'} \chi_{sp_y}(-1)^{i_y+i'_y} S ^{sp_y}_{i,x}S ^{sp_y}_{i^\prime,x} + \sum_{i}\chi_{p_xp_ysp_xsp_y}(-1)^{i_x+i_y} S ^{p_xp_y}_{i,x}\nonumber\\
	&+&  \sum_{ii'}(\chi_{p_xp_ysp_x}+\chi_{p_xp_ysp_y})(-1)^{i_x+i_y + i'_x+i'_y} S ^{p_xp_y}_{i,x}
	S ^{p_xp_y}_{i^{\prime},x}\nonumber\\
	&-& \frac{1}{2}\sum_{i} \left(       \frac{\mu_{s} -\mu_{p_x} }{2}S ^{sp_x}_{i,z} +  \frac{\mu_{s} -\mu_{p_y} }{2}S ^{sp_y}_{i,z} +\frac{\mu_{p_x} -\mu_{p_y} }{2}S ^{p_xp_y}_{i,z}\right) + H_{sp_x} + H_{sp_y} + H_{p_xp_y}, 
\end{eqnarray}
where
\begin{eqnarray}
	\chi_{sp_x}   &=&  \left( 8\mathrm{Re}[ \tilde{\Delta}_{2}/\mathcal{Z}] - |2\tilde{\Delta}_{2} /\mathcal{Z} |^2    \Delta_1 \right) J^{ii}_{sp_x} J^{i^\prime i^\prime}_{sp_x}  ,
\end{eqnarray}
\begin{eqnarray}
	\chi_{sp_y}   &=&  \left( 8\mathrm{Re}[ \tilde{\Delta}_{1}/\mathcal{Z}] - |2\tilde{\Delta}_{1} /\mathcal{Z}|^2 \Delta_2 \right)  J^{ii}_{sp_y} J^{i^\prime i^\prime}_{sp_y} ,
\end{eqnarray}

\begin{eqnarray}
	\chi_{p_xp_ysp_xsp_y }   &=&  \left\lbrace \mathrm{Re}[32 /\mathcal{Z}] +  16( \mathrm{Re}[\tilde{\Delta}_{1} /\mathcal{Z}] \mathrm{Re}[\tilde{\Delta}_{2} /\mathcal{Z}]+ \mathrm{Im}[\tilde{\Delta}_{1} /\mathcal{Z}]\mathrm{Im}[\tilde{\Delta}_{2} /\mathcal{Z}] ) \right.\nonumber\\ &-& 2 \Delta_1 (\mathrm{Re}[4/\mathcal{Z}]\mathrm{Re}[2\tilde{\Delta}_{2}/\mathcal{Z}]  +\mathrm{Im}[4/\mathcal{Z}]\mathrm{Re}[2\tilde{\Delta}_{2}/\mathcal{Z}])    \nonumber\\ &-& \left.2 \Delta_2 (\mathrm{Re}[4/\mathcal{Z}]\mathrm{Re}[2\tilde{\Delta}_{1}/\mathcal{Z}]  +\mathrm{Im}[4/\mathcal{Z}]\mathrm{Re}[2\tilde{\Delta}_{1}/\mathcal{Z}])    \right\rbrace \sum_{i^{\prime},i^{\prime\prime}}(-1)^{i^\prime_x+i^{\prime \prime}_y} J^{i i}_{p_{x}p_{y}}J^{i^\prime i^\prime}_{sp_{x}}  J^{i^{\prime \prime}i^{\prime \prime}}_{sp_y}S ^{sp_x}_{i^\prime,x}  S ^{sp_y}_{i^{\prime \prime} ,x}  ,
\end{eqnarray}
\begin{eqnarray}
	\chi_{p_xp_ysp_x}   &=& \left\lbrace  4( \mathrm{Re}[2\tilde{\Delta}_{2} /\mathcal{Z}] \mathrm{Re}[4 /\mathcal{Z}] + \mathrm{Im}[2\tilde{\Delta}_{2} /\mathcal{Z}]\mathrm{Im}[4 /\mathcal{Z}]) - |4/\mathcal{Z}|^2\Delta_2 \right \rbrace \nonumber \\ 
	&\times& \sum_{i^{\prime\prime},i^{\prime\prime\prime}} (-1)^{i^{\prime\prime}_x+i^{\prime\prime\prime}_x} J^{i^{\prime\prime} i^{\prime\prime}}_{sp_x}J^{i^{\prime\prime\prime}i^{\prime\prime\prime}}_{sp_x}J^{i i}_{p_{x}p_{y}} J^{i^{\prime}i^{\prime}}_{p_xp_y} S ^{sp_x}_{i^{\prime\prime},x} S ^{sp_x}_{i^{\prime\prime\prime},x} ,
\end{eqnarray}
\begin{eqnarray} 
	\chi_{p_xp_ysp_y}   &=&  \left\lbrace 4( \mathrm{Re}[4 /\mathcal{Z}] \mathrm{Re}[2\tilde{\Delta}_{1} /\mathcal{Z}] + \mathrm{Im}[4 /\mathcal{Z}]\mathrm{Im}[2\tilde{\Delta}_{1} /\mathcal{Z}]) - |4/\mathcal{Z}|^2 \Delta_1  \right \rbrace \nonumber \\ 
	&\times& \sum_{i^{\prime\prime},i^{\prime\prime\prime}} (-1)^{i^{\prime\prime}_y+i^{\prime\prime\prime}_y} J^{i^{\prime\prime} i^{\prime\prime}}_{sp_y}J^{i^{\prime\prime\prime}i^{\prime\prime\prime}}_{sp_y}J^{i i}_{p_{x}p_{y}} J^{i^{\prime}i^{\prime}}_{p_xp_y} S ^{sp_y}_{i^{\prime\prime},x} S ^{sp_y}_{i^{\prime\prime\prime},x},
\end{eqnarray}

\begin{eqnarray}
	H_{sp_x} &=&\sum_i  (\frac{U_s +U_{p_{x}} }{16 } - \frac{  U_{sp_{x}}}{ 2 }  ) {S ^{sp_x}_{i,z}}^2 + U_{sp_{x}}\left({S ^{sp_x}_{i,x}}^2-{S ^{sp_x}_{i,y}}^2\right) +\frac{U_s-U_{p_{x}} }{8}( n_{i,sx}-1)  S ^{sp_x}_{i,z} \nonumber \\ &+
	& (\frac{U_s +U_{p_{x}} }{16} + \frac{U_{sp_{x}}}{2})n_{i,sx}^2 - \frac{U_s + U_{p_{x}}}{8} n_{i,sx} , 
\end{eqnarray}

\begin{eqnarray}
	H_{sp_y}&=&\sum_i  (\frac{U_s +U_{p_{y}} }{16 } - \frac{  U_{sp_{y}}}{ 2 }  ) {S ^{sp_y}_{i,z}}^2 + U_{sp_{y}}\left({S ^{sp_y}_{i,x}}^2-{S ^{sp_y}_{i,y}}^2\right) +\frac{U_s-U_{p_{y}} }{8}( n_{i,sy}-1)  S ^{sp_y}_{i,z} \nonumber \\ &+
	& (\frac{U_s +U_{p_{y}} }{16} + \frac{U_{sp_{y}}}{2})n_{i,sy}^2 - \frac{U_s + U_{p_{y}}}{8} n_{i,sy} ,
\end{eqnarray}
\begin{eqnarray}
	H_{p_xp_y}&=&\sum_i  (\frac{U_{p_x} +U_{p_{y}} }{16 } - \frac{  U_{p_xp_{y}}}{ 2 }  ) {S ^{p_xp_y}_{i,z}}^2 + {U_{p_xp_y}}\left({S ^{p_xp_y}_{i,x}}^2-{S ^{p_xp_y}_{i,y}}^2\right) +\frac{U_{p_x}-U_{p_{y}} }{8}( n_{i,xy}-1)  S ^{p_xp_y}_{i,z} \nonumber \\ &+
	& (\frac{U_{p_x} +U_{p_{y}} }{16} + \frac{U_{{p_x}p_{y}}}{2})n_{i,xy}^2 - \frac{U_{p_x} + U_{p_{y}}}{8} n_{i,xy} ,
\end{eqnarray}
and $n_{i,sx} = n_{i,s} + n_{i,p_x}$, $n_{i,sy} = n_{i,s} + n_{i,p_y}$, $n_{i,xy} = n_{i,p_x} + n_{i,p_y}$.

\end{widetext}

\bibliography{references}

\end{document}